# Implementation of Uplink Network Coded Modulation for Two-Hop Networks

**Yi Chu, Tong Peng, David Grace and Alister Burr**

University of York,York, YO10 5DD, UK

Corresponding author: Yi Chu (e-mail: yi.chu@york.ac.uk).

The presented research work is supported by the EPSRC grant *Network Coded Modulation for Next Generation Wireless Access Networks* with reference EP/K040006/1. The work is also supported by the project partners BT, Fujitsu and Vodafone. National Instruments has provided technical support to the software and hardware used in the experiments.

**ABSTRACT** With the fast growing number of wireless devices and demand of user data, the backhaul load becomes a bottleneck in wireless networks. Physical layer network coding (PNC) allows Access Points (APs) to relay compressed, network coded user data, therefore reducing the backhaul traffic. In this paper, an implementation of uplink Network Coded Modulation (NetCoM) with PNC is presented. A 5-node prototype NetCoM system is established using Universal Software Radio Peripherals (USRPs) and a practical PNC scheme designed for binary systems is utilised. An orthogonal frequency division multiplexing (OFDM) waveform implementation and the practical challenges (e.g. device synchronisation and clock drift) of applying OFDM to NetCoM are discussed. To the best of our knowledge this is the first PNC implementation in an uplink scenario in radio access networks and our prototype provides an industrially-applicable implementation of the proposed NetCoM with PNC approach.

## I. INTRODUCTION

Terrestrial wireless networks are now widely used and serve the vast majority of the world's population. This has resulted in an ever-increasing volume of traffic, which is being met in urban areas by densely deployed Access Points (APs). These densely deployed infrastructures result in a need for a lower cost per node, which means that more processing is moved from the edge of the network into the wider cloud. Access points will act as smart relays rather than fully process the data traffic. These dense deployments also result in the potential for significant coverage overlap especially at cell-edge, which could result in unwanted interference or instead an opportunity to exploit joint transmission or reception, allowing for greater exploitation of the fading environment.

### A. *Motivations*

Coordinated Multipoint (CoMP) has been introduced in LTE-A Release 11 [1] to provide better Quality of Service (QoS) to cell edge users using joint reception on uplink, where the users with weak signal strength can be served by multiple APs simultaneously to improve channel performance. However, these benefits come with costs, CoMP trades backhaul traffic load with access link performance and the additional backhaul load depends on the quantization scheme, co-channel interference, modulation and the number of APs involved in joint reception. The bottleneck on the backhaul can significantly affect the network throughput, particularly when the backhaul uses wireless links which have limited capacity. Physical Layer Network Coding (PNC) is a technique which allows multiple sources to transmit on the same carrier frequency [2], with the user data being restored at the hub. NetCoM uses PNC to convert what would be interfering signals between users into useful data streams. Each AP uses PNC to encode the superimposed analogue user signals to coded bit streams which have the same length as the original data and forwards them to the hub, thereby reducing the backhaul traffic load. Moreover, NetCoM can also be applied to non-cell edge users to improve the spectral efficiency on access links if the throughput bottleneck appears on the access links.

### B. *Related Work*

The term network coding (NC) was first presented in [3] which introduces the possibility of coding on layer-3 packets to replace direct packet forwarding, thereby improving the overall network throughput. The authors of [2] extend the idea of NC to wireless communications by exploiting the additive nature of simultaneously arriving electromagnetic (EM) waves as coding "in the air", where in a two-way-relay (TWR) network the throughput can in principle be doubled. PNC in TWR networks has been extensively studied over the past few years. Thorough theoretical performance analysis of PNC on TWR networks has been presented in [4] with closed-form expressions for the outage probability, with performance compared to the original NC. Another performance analysis of PNC has been presented in [5] including the effect of channel estimation errors and frequency-selective fading. Errors in fading tracking cause more performance degradation than channel noise and sophisticated channel estimation techniques are required for selective fading channels. A mapping and constellation design algorithm for TWR PNC has been proposed in [6]. The design criterion focuses on non-conventional constellations other than traditional m-array phase shift keying (M-PSK) where different constellations are used at all three nodes in the network. Two different design methods have been presented to study the trade-off between throughput performance and computational complexity. Results show that they both outperform the benchmark scheme.

The issue of source signal asynchrony has been studied in [7], where the authors propose a solution for decoding at the receiver based on belief propagation and demonstrate the performance for QPSK modulation for both coded and uncoded systems. For a similar purpose the authors of [8] have proposed a synchronisation scheme for PNC with QAM modulation schemes. Results in this paper show that PNC outperforms NC even after including the effects of synchronisation errors and overheads. Challenges and solutions for designing PNC for fading TWR networks have been discussed in [9], and a new PNC design criterion has been proposed to produce optimised integer coefficients that minimise the error probability of PNC operating with Rayleigh fading channels. The impact of relative phase offset of the signals received at the relay node has been investigated in [10]. The channel decoder performance degradation is studied and a symbol-level PNC decoder is proposed with improved performance over bit-level decoder. The decoder performance is demonstrated via both simulations and experiments.

The engineering applicability of PNC to TWR networks has been demonstrated by prototypes developed by many researchers. Software Defined Radio (SDR) tools have been widely used to evaluate the techniques developed for PNC in the real world. TWR PNC has been implemented on USRP platforms using the OFDM waveform in [11][12], and the authors have identified many issues while applying PNC to practical systems. This prototype performs XOR mapping in the frequency domain by using symbols carried by each subcarrier rather than time domain samples. The results show similar performance with symbol-synchronous and symbol-asynchronous conditions. It is also possible to apply PNC to asynchronous TWR networks [13][14] and USRP testbeds are used to evaluate the system performance. A convolutional-coded PNC scheme has been proposed in [13] to compensate carrier phase offset and symbol misalignment. An asynchronous discrete-time model has been proposed in [14] which transfers device impairments between relay and users to ensure data recovery. A carrier frequency offset (CFO) correction scheme has been introduced in [15] to compensate the CFO between source nodes and relay, and the system is evaluated using SDR testbeds. The authors of [16] have integrated PNC to TWR networks with 802.11 MAC/PHY layers. Both symmetric relaying (same source modulation scheme) and asymmetric relaying (different source modulation schemes) have been studied and evaluated using SDR testbeds.

The work presented in [17][18][19] has extended the SDR implementation to two-hop networks. The authors of [17] have presented a nonlinear fixed-gain amplify-and-forward relay system and evaluated the system with OFDM waveform using SDR testbeds. The symbol error rate (SER) of the practical system has been compared with the theoretical benchmark to demonstrate the robustness of the system. A distributed learning algorithm for PNC mapping selection has been proposed in [18] and demonstrated by a practical system. The learning algorithm determines the optimal mapping at the relay nodes and ensures invertibility at the destination nodes by using the information exchanged between relays. The results of the experiments demonstrate the convergence of the learning algorithm. The authors of [19] have proposed the concept of wireless cloud networks which uses many relay nodes capable of PNC to handle the traffic between source and destination nodes. A cloud initialisation procedure is presented to estimate the number of sources and the associated channel parameters based on the received superimposed constellations. The superimposed constellations at relay nodes are demonstrated by a practical system using BPSK modulation. Note that the above work all requires common time and frequency references to be distributed to SDR testbeds.

C. *Contributions*

The existing PNC work in TWR networks can benefit point-to-point (2 UEs and a relay node) transmissions with symmetric traffic levels. However users usually have significantly different traffic requirements on the uplink and downlink. For example watching a 4K film online requires fast download speed and live streaming needs good uplink throughput. In this paper, we apply PNC to the uplink of a two-hop network designed to model a radio access network (RAN) in which two user equipment (UE) nodes, two APs and one hub form one 5-node interference coordination group (ICG).

Fig. 1 shows an example RAN network with 5 APs connected to a Hub via wireless backhaul. All UEs within the range of two or more APs can potentially be involved in one (or more) ICG. For example UE9 and UE10 are within the range of AP3 and AP5, these 4 nodes and the Hub can potentially form an ICG. Conventionally UE9 and UE10 should use orthogonal access link radio resource (either in time or frequency) to avoid interference, however PNC allows both UEs to be served by two APs on the same resource, thereby improving the access link spectral efficiency. The number of ICGs benefits significantly from the overlapped cells. For example UE7 is within the range of AP3, AP4 and AP5, and it can form ICGs with UE6, UE8, UE9 or UE10. It can be noted that in future networks with vastly increasing UEs and densely deployed APs, there will be many opportunities to form ICGs which are capable of utilising interference sources to useful signals.

In this paper we demonstrate the physical layer signal processing of each 5-node ICG using real-world experiments. To the best of our knowledge this is the first time that PNC is implemented in a RAN architecture in the uplink scenario using SDR testbeds. This prototype system demonstrates that NetCoM could be a viable operational interference coordination technique and an alternative to CoMP.

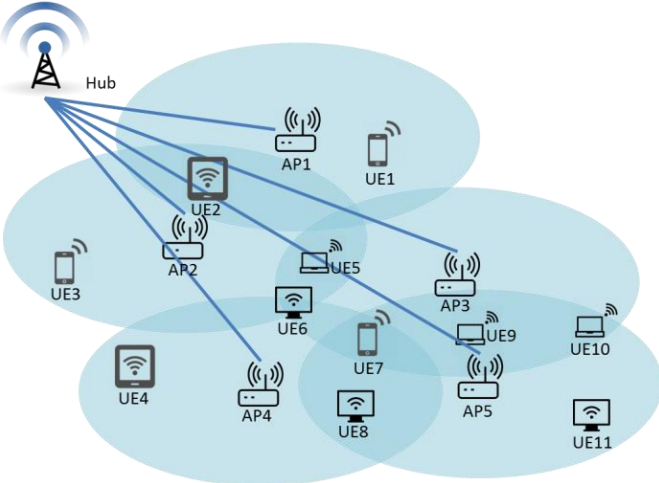

Fig. 1 PNC in a RAN network

The rest of the paper is organised as follows. Section II describes the NetCoM system design and introduces the offline and online PNC mapping search to address singular fading states. Section III discusses the practical challenges of SDR implementation and the resulting solutions. Section IV presents the implementation scenario and the system configurations. Section V illustrates the performance of the system under different fading situations. Section VI concludes the paper.

## II. System Design

This section introduces the 5-node NetCoM system model evaluated by SDR testbeds and offline/online PNC mapping search.

### A. System Model

Fig. 2 presents the 5-node ICG NetCoM system model for uplink networks. The wireless links from UEs to APs represent access links and the links from APs to hub are backhaul links. Each AP receives 4-QAM modulated signals from both UEs and estimates the relative channel state information (RCSI) by using pilots. Both APs send the RCSI to the hub for it to determine the appropriate PNC mapping. After receiving the mapping the APs forward linear combinations (referred to in this paper as the network coded symbols) of UE messages over a finite field to the hub to recover the UE data [20]. We assume that all UEs and APs use a single antenna. The backhaul links can be wireless or wireline. A NetCoM system can particularly benefit wireless backhaul which is usually more bandwidth limited. The system can be extended to beyond 5 nodes and higher order modulation schemes at the cost of more computational complexity, however in this paper we will focus on demonstrating the 5-node and 4-QAM case evaluated by SDR.

The noisy, faded and superimposed signal received by the $j^{th}(j = 1, 2)$ AP can be represented by:

$$y_j = h_{j1}s_1 + h_{j2}s_2 + z_j \quad (1)$$

where $z_j$ is a complex Gaussian random variable with zero mean and $\sigma^2$ variance per dimension, $h_{ji}$ represents a channel coefficient between the $i^{th}(i = 1, 2)$ UE and the $j^{th}$ AP, and $s_i$ is the 4-QAM modulated complex symbol sent by the $i^{th}$ UE.

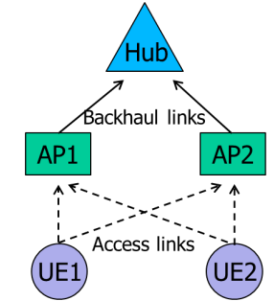

Fig. 2 5-node NetCoM system

### B. PNC Design

PNC compresses the messages from all UEs by remapping the received signals to network codewords using algebraic approaches to reduce the cardinality of the output from all APs while maintaining the overall throughput, thereby reducing the backhaul traffic loads and improving the network throughput. UE messages are restored at the hub by using two NCS streams from APs. Three issues need to be resolved for PNC to be applicable:
1) The PNC decoding operates over the binary field $\mathbb{F}_2$, thus the NCS must be binary based.
2) The PNC mapping function must be well designed to resolve all singular fade states (SFS).
3) The hub must be able to restore both UE messages based on the two NCS streams from both APs.

Instead of using the PNC approach in [21] where the linear combinations are performed on complex symbols, we design the PNC mapping directly on the message bits, thus the AP decodes the linear network coded function (LNCF) over $\mathbb{F}_2$. We define $\mathcal{N}_j$ as the bit-level LNCF of the $j^{th}$ AP, $\mathbf{x}_j$ as the network coded vector (NCV) consisting 2 linear network coded bits (LNCB):

$$\mathbf{x}_j = \mathcal{N}_j(\mathbf{M}_j, \mathbf{w}) = \mathbf{M}_j \star \mathbf{w} \quad (2)$$

$$\mathbf{x}_j = \left[x_j^{(1)}, x_j^{(2)}\right]^T \quad (3)$$

where $\mathbf{w}$ denotes the $1 \times 4$ joint UE message binary vector, $\mathbf{M}_j$ denotes the $2 \times 4$ binary mapping matrix and $\star$ denotes the modulo-2 matrix multiplication. The hub restores the UE message when the NCVs and mapping matrices of both APs are known:

$$\mathbf{w} = \begin{bmatrix}\mathbf{M}_1\\\mathbf{M}_2\end{bmatrix}^{-1} \star \begin{bmatrix}\mathbf{x}_1\\\mathbf{x}_2\end{bmatrix} \quad (4)$$

Let $\mathbf{s} \triangleq [s_1, s_2]$ denote the vector containing superimposed modulated symbols from both UEs: we define all the possible superimposed constellation points at the $j^{th}$ AP with channel coefficient vector $\mathbf{h}_j \triangleq [h_{j1}, h_{j2}]$ as a vector given by

$$\mathbf{s}_{j,\triangle} \triangleq \left[s_{j,\triangle}^1, \cdots, s_{j,\triangle}^{16}\right] = h_{j1}s_1 + h_{j2}s_2 \quad (5)$$

where $s_{j,\triangle}^{(\tau)}$ denotes the superimposed constellation point arising from two 4-QAM modulated symbols $s_1$ and $s_2$, and $\tau = 1, 2, \cdots, 16$ denotes the index of all possible combinations of two

4-QAM modulated symbols. For particular channel coefficient vectors, some superimposed symbols have the same value, then the SFS is defined as the channel coefficient $\mathbf{h}_j$ which makes $s_{j,\triangle}^{(\tau)} = s_{j,\triangle}^{(\tau')}$ for some $\tau \neq \tau'$. For example, Fig. 3 shows the superimposed constellation points when $h_{j1} \approx h_{j2} \approx 1$: the points corresponding to source bits (00, 01) and (01, 00) are very close to each other, due to singular fading.

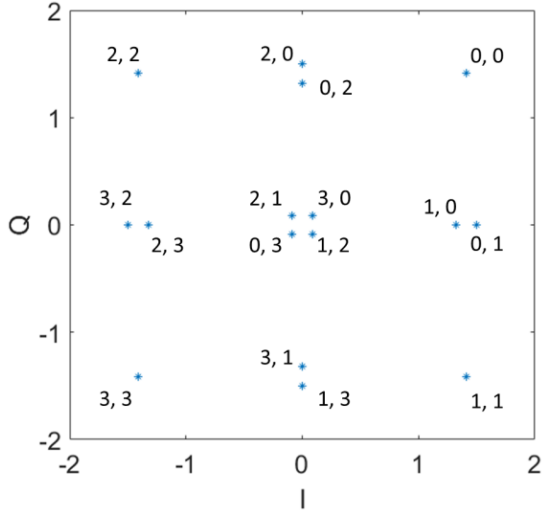

**Fig. 3 Superimposed constellation**

Normally, singular fading is unavoidable in the multiple access channel and multiuser detection is in principle infeasible if the $j^{th}$ AP expects to decode all source messages. PNC provides a good solution to overcome SFS by ensuring that the coincident superimposed constellation points are labelled by the same NCV $\mathbf{x}_j$ which helps the hub to recover all source bits. The PNC mapping is defined by a surjective function:

$$\Theta: \mathbf{s}_{j,\triangle} \longrightarrow \mathbf{x}_j \tag{6}$$

and the minimum distance between NCVs is defined as:

$$d_{min} = \min_{\forall \tau,\tau': \Theta(s_{j,\triangle}^{(\tau)}) \neq \Theta(s_{j,\triangle}^{(\tau')})} \left| s_{j,\triangle}^{(\tau)} - s_{j,\triangle}^{(\tau')} \right|^2 \tag{7}$$

Thus, if all superimposed constellation points that can be labelled by the same NCV are placed in a certain cluster, $d_{min}$ is the minimum inter-cluster distance.

### C. Binary Mapping Matrix Design

The PNC mapping function should be designed with the following properties:

1) To keep the maximum distance between the NCVs.
2) To ensure that the combined mapping matrix is invertible.

To achieve these binary matrices an adaptive selection algorithm with two steps is proposed. The first is an Off-line search, which implements an exhaustive search among all $\mathbf{M}_j$s to find a set of candidate matrices which resolves all SFSs with property 1). The second step is an On-line search, which selects the combined mapping matrix according to property 2). The computational complexity is dominated by the Off-line search, particularly for higher order modulation schemes. However, this only needs to be performed once and all candidate mapping matrices are then stored at APs and hub for use in the On-line search.

We define $\mathbf{v}_{SFS} = [v_{SFS}^{(1)}, \cdots, v_{SFS}^{(l)}]$ as a vector that contains all SFSs and each element is calculated by:

$$v_{SFS}^{(l)} = \frac{h_{j2}}{h_{j1}} = \frac{s_1^{(\tau)} - s_1^{(\tau')}}{s_2^{(\tau')} - s_2^{(\tau)}} \tag{8}$$

Then when an SFS occurs,

$$h_{j1} s_1^{(\tau)} + h_{j2} s_2^{(\tau)} = h_{j1} s_1^{(\tau')} + h_{j2} s_2^{(\tau')} \tag{9}$$

where $s_\ell^{(\tau)}$ and $s_\ell^{(\tau')}$ represent different modulated symbols at the $\ell^{th}$ UE. The constellation points $s_{n,\triangle}^{(l)}$ and corresponding NCV $\mathbf{x}_{i,n}$ under all $L$ SFS can be given as:

$$s_{n,\triangle}^{(l)} = \mathbf{v}_{SFS}^{(l)} \mathbf{s}_n, \ n = 1, \cdots, N, \ l = 1, \cdots, L \tag{10}$$

$$\mathbf{x}_{i,n} = \mathbf{M}_i \star \mathbf{w}_n, i = 1, \cdots, N^2 \tag{11}$$

where $\mathbf{v}_{SFS}^{(l)} = [v_{SFS}^{(l)} \ 1]$ denotes the channel coefficient vector, $\mathbf{s}_n$ is the $n^{th}$ joint modulated symbol vector, $\mathbf{w}_n$ is the $n^{th}$ joint source message vector, and $N$ is the number of different combinations of source bits from all UEs. In the case of 4-QAM modulation, $N = 16$.

When an SFS occurs, different $\mathbf{s}_n$ can be mapped to the same constellation point $s_{n,\triangle}^{(l)}$, and these $\mathbf{s}_n$ should be encoded with the same PNC codeword so that the hub is able to restore the UE messages. A vector $\mathbf{Q}_d$ can be defined to contain all $d_{min}$ between different NCS vectors for every binary mapping matrix in each SFS. A table containing the minimum distance $d_{min}$, constellation points $s_{n,\triangle}^{(l)}$, PNC codeword $\mathbf{x}_{i,n}$, joint message vector $\mathbf{s}_n$ and mapping matrix $\mathbf{M}_i$ can be established. All image SFSs (that is, SFSs that can be resolved by the same mapping) are removed to reduce the number of candidate mapping matrices. The mapping matrices which cannot form a full rank matrix with other candidate matrices are also removed.

The successful candidate matrices from the Off-line search are stored at the APs and hub for use in the On-line search to select the optimal mapping. A maximum $d_{min}$ should be achieved between NCS clusters and the combined matrix should be full rank to ensure decodability at the hub. Before the Off-line search, the values of SFSs can be calculated according to the constellations used at each UE and stored at APs and the hub, e.g. 5 SFSs for 4-QAM used at both UEs. After receiving the superimposed UE signals, APs use the estimated channel coefficients to calculate the closest SFS and these SFS indices are sent to the hub. Using the stored mapping matrix candidates obtained from the Off-line search, the optimal combined mapping matrix, which returns the largest minimum distance between the NCVs, can be selected by the hub which sends the mapping index to both APs. The APs then try to decode both UE message bits according to the channel coefficients and encode the message bits to PNC codewords by performing (2) using the optimal mapping. The same PNC codewords can be obtained from the combinations of message bits mapped to the same NCV via the modulo-2 matrix multiplication. UE messages are restored at the hub by multiplying the inverse mapping matrix with the combined NCV from both APs.

For the 4-QAM case demonstrated in the prototype, five SFSs have been found and the relative channel coefficients $h_{j2}/h_{j1}$ of these SFSs are used to determine which SFS the current signal is closest to, thereby determining the optimal PNC mapping the APs should use. The Off-line search finds five optimal mapping matrices to resolve the SFSs which lead to 25 possible mapping combinations. For each combination of SFSs from the two UEs, the optimal full-rank mapping matrix, which maximises the minimum distance between NCVs, is selected from these 25 matrices in the On-line search. The SFSs and mappings are included in the appendix.

### III. Practical Challenges and Solutions

This section lists the practical challenges for prototyping the NetCoM system and presents our solutions.

#### A. Implementation Overview

To practically implement the system presented in Fig. 2, multiple SDR devices are required to operate as different nodes in the network. In the prototype system we use four USRP devices to operate as five nodes, where the functions of two UE nodes are performed by a single USRP. All packet exchanges between USRP devices are wireless, where the access links are multiple access channels and the backhaul links are delivered using orthogonal channels. RCSI is critical for PNC to determine the correct mapping to encode the UE messages. Failure to obtain the accurate RCSI could affect the system performance significantly. Frequency and time asynchrony naturally exist between USRP devices because of the clock drift of the oscillator and they all affect the RCSI estimation if not compensated correctly. The issues of asynchrony are discussed in later sections along with solutions.

#### B. Device Synchronisation

In the prototype system, all access link and backhaul link signals are modulated by OFDM. The USRP-2943R SDR testbed is used and its internal oscillator has an accuracy of 2.5 ppm [22]. Each USRP has two daughter boards with two antenna ports, and the daughter boards generate their carrier frequencies independently using their internal oscillator. Different rates of the oscillators cause CFO to the OFDM signals, which needs to be compensated. In our system two daughter boards are used on one USRP to transmit the two UE signals simultaneously. Three other USRPs are used as two APs and one hub node as Fig. 4 shows. In this case, CFO exists in the user signals when they arrive at both APs, and it also exists in all signals between two APs and the hub. The CFO estimation and correction of IEEE 802.11a [23] is applied using preambles on all OFDM signals to prevent inter-carrier-interference (ICI).

For the receiver to detect the OFDM signal, a pseudo-noise (PN) sequence is applied at the head of each OFDM frame so that the receiver can determine the start of an OFDM frame by calculating the correlation. This PN sequence is sent by one UE while the other UE remains silent. A cyclic prefix (CP) is applied to remove the effect of time offsets caused by the delay between two UE signals, or minor frame detection errors from the PN sequence. The time offset between UE signals could be positive or negative, so CP is applied at both ends of the OFDM symbol to ensure the data integrity. Fig. 5 explains the reason why the CP is applied at both ends of the data symbol. The starting sample of each OFDM frame is determined by the correlation of the PN sequence sent by UE A, and the AP removes the CP according to the detection result. The CP placed at the head of each data symbol can protect the data if the signal of UE B is later (within the CP) than the signal of UE A. However, if the signal of UE B arrives earlier than UE A as Fig. 5(a) shows, part of the data symbol of UE B will be lost after removing the CP. By applying the CP at both ends of the data symbol as Fig. 5(b) shows, the data will not be lost in either delay condition.

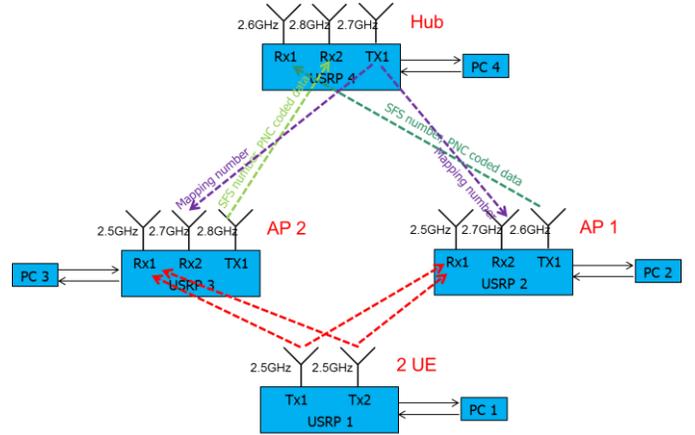

Fig. 4 Experiment scenario

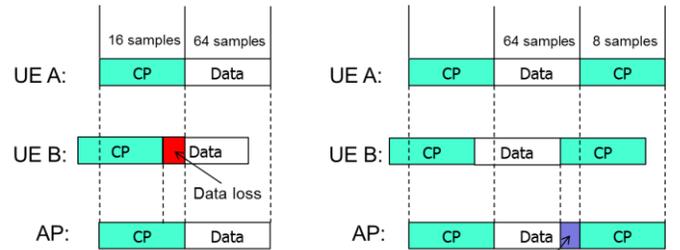

Fig. 5(a) CP at the head   (b) CP at both ends

#### C. RCSI Estimation

The appropriate PNC mapping is selected based on the RCSI estimated at both APs. Channel estimation is performed at each AP for both UE signals by using pilots transmitted on separate subcarriers. Initially two pilot carriers were used in every six subcarriers for demonstration purposes, but it is anticipated that the number of pilot carriers can be further reduced in the future. In this case, out of every six subcarriers of each UE signal, there are four data carriers, one pilot carrier and one null carrier where the null carrier is reserved as a pilot carrier of the other UE. When both channels are known at the AP, the AP determines which SFS the signal is closest to (according to the distance between the estimated RCSI and the SFSs provided in the appendix) and sends the SFS index to the hub. With SFS indices of both APs known at the hub, the hub sends back the appropriate mapping index for the APs to encode UE data using PNC.

In our prototype system the start time for each USRP to transmit or receive is determined by a software trigger from the host PC. All USRP devices are controlled by individual host PCs so the trigger time is not synchronised. The random trigger time causes sampling clock offset (SCO) between transmitting and receiving devices. This offset can be observed on most received signals unless in rare cases the sampling clock of both devices is aligned by coincidence. Clock drift across USRP devices is another source of SCO, a periodically changing subcarrier phase difference can be observed on continuous transmission and reception. With the frame detection of PN sequence, a residual delay $\Delta t$ still exists where $0 \leq \Delta t \leq t_s/2$ and $t_s$ is the sample duration. Given an OFDM signal with delay $\Delta t$

$$S(t - \Delta t) = \frac{1}{\sqrt{N}} \sum_{k=0}^{F-1} x(k) e^{j\frac{2\pi k(t-\Delta t)}{T}} \qquad 0 \leq t \leq T \quad (12)$$

where $x(k)$ is the data sequence, $F$ is the FFT length, $T$ is OFDM symbol duration (excluding CP) and $T = F t_s$. A phase offset $\varphi$ can be observed between neighbouring subcarriers

$$\varphi = e^{j\frac{2\pi(-\Delta t)}{T}} \qquad (13)$$

and it causes a linear phase shift across successive subcarriers. In every OFDM symbol, $\varphi$ can be obtained from the phases of pilot subcarriers and $\varphi$ is used with the channel estimated on pilot subcarriers to estimate the channels on data subcarriers.

### D. Packet Exchange between Two APs and Hub

To complete one PNC encoded data packet transmission (from each AP), three packets need to be exchanged on the backhaul links between the hub and each AP. Both APs first send the SFS indices to the hub, wait for hub to reply with the PNC mapping index, then send the PNC encoded data to the hub. A basic MAC layer is designed to control the packet delivery on backhaul link.

In the prototype system, all baseband signal processing is completed by the host PC. For each received sample stream on the USRP, approximately 15,000 samples can be stored in the memory and the samples in the memory will be replaced by new samples if the PC cannot process the samples faster than the USRP generates them. The sample rate (IQ rate) that was used is 1M samples/s on all USRP devices. Due to hardware limitations on the type of USRP boards used and the speed of processing on the PC, the memory on every USRP always becomes full after the system starts. This means samples are constantly discarded because of the relatively longer processing time of the PC. To overcome this situation in our experiments the APs and hub always send multiple copies of the same packet to make sure at least one of them is processed by the receiver. Fig. 6(a) and (b) show the flow chart of both APs and the hub. A timeout limit of one second is given to mapping index reception at each AP, so that if the AP has not received mapping index from hub within one second after it sends the SFS index the AP will attempt to encode the UE data using previously obtained PNC mapping.

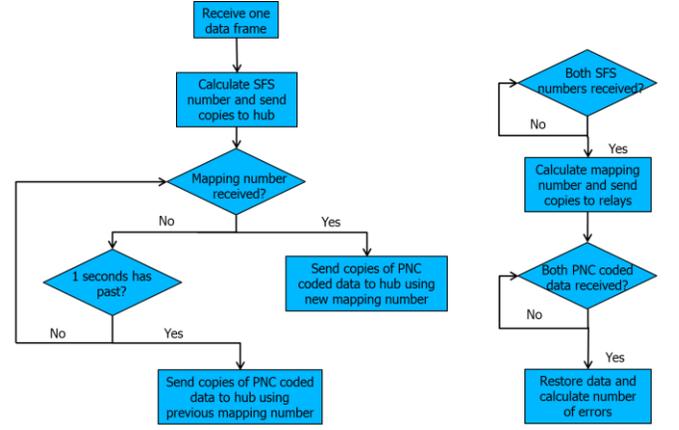

Fig. 6(a) AP flow graph      (b) Hub flow graph

### E. UE Data Frame Structure

The two APs use preambles sent by UE A for frame detection and CFO correction, while UE B only sends data symbols simultaneously with data symbols from UE A. Fig. 7 shows the frame structure of both UEs. Three preambles are transmitted by UE A for frame detection, coarse CFO correction and fine CFO correction respectively. The FFT length is 64 and CP length is 16 samples. Note that all preambles use the conventional CP because the issue described in Fig. 5 will not occur for preamble reception.

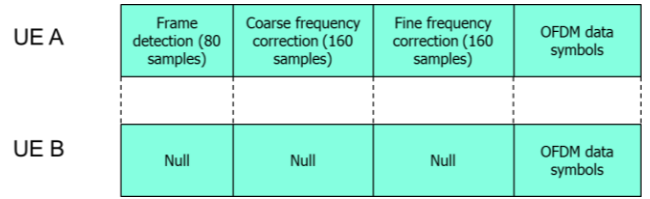

Fig. 7 UE frame structure

## IV. Experiment Setup

This section describes the experimental scenario and system parameters of the prototype system.

### A. Experiment Hardware

Four USRP-2943R are used in this system. Three USRPs are connected to the PCIe slots on three desktop PCs, the last USRP (serving two-UE nodes) is connected to a laptop via a Sonnet Echo Express SE I thunderbolt to PCIe adapter [24] to improve the mobility of this node. We use Log Periodic antennas [25] which cover 850 MHz to 6.5 GHz and allow signals to be sent over a wide range of carrier frequencies. In-house radio anechoic chamber measurements show that their 3dB beamwidth is 200° with maximum antenna gain 5 dBi, with approximately 20 dB cross polar attenuation. This antenna directivity and polarisation mean that the antennas must be correctly aimed at one another during the experiments to ensure that the two UE signals have similar signal power levels when they arrive at each AP.

### B. Experiment Scenario

Frequency Division Duplexing (FDD) is applied to place the access links and backhaul links on separate carrier frequencies to avoid internal interference within the system. As Fig. 4 shows the access links from two UEs to two APs are assigned 2.5 GHz. The uplink backhaul from APs 1 to the Hub uses 2.6 GHz, while the other uplink backhaul from AP 2 to the Hub uses 2.8 GHz. The downlink backhaul from Hub to both APs uses 2.7 GHz. Fig. 8 shows a picture of the system in an indoor lab. Similarly to Fig. 4, the nodes and links are marked in the picture. Wooden burette stands are used to hold the antennas and maintain their polarization. The stands also allow the antennas to rotate to different directions to ensure the received signal strength can be maximized, particularly for the superimposed UE signals.

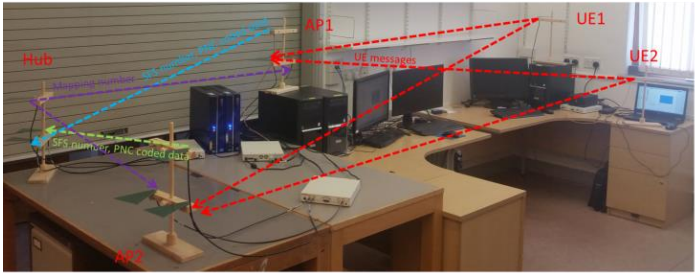

**Fig. 8 Experiment scenario**

### C. System Parameters

Table I shows the system parameters used during the experiments. The system is initiated by APs receiving superimposed packets from two UE. The UE packets are sent in bursts, with each burst containing 10 packets. The time between bursts is configurable. Note that the listed lengths of packets are for evaluation purposes only rather than to maximize performance. The payload length in each packet can be potentially extended taking into account the maximum clock drift between USRP devices.

TABLE I
SYSTEM PARAMETERS

| Parameters | Values |
|---|---|
| Access link carrier frequency | 2.5 GHz |
| AP 1 uplink backhaul carrier frequency | 2.6 GHz |
| AP 2 uplink backhaul carrier frequency | 2.8 GHz |
| Downlink backhaul carrier frequency | 2.7 GHz |
| Baseband sample rate (IQ rate) | 1 M samples/s |
| Signal bandwidth | 1 MHz |
| FFT length | 64 |
| CP length | 16 |
| Number of used subcarriers | 48 |
| Subcarrier spacing | 15.625 kHz |
| UE packet length | 880 samples |
| AP packet length (SFS index) | 560 samples |
| AP packet length (PNC encoded UE data) | 800 samples |
| Hub packet length (mapping index) | 560 samples |
| Access link carrier frequency | 2.5 GHz |
| AP 1 uplink backhaul carrier frequency | 2.6 GHz |

## V. Results and Discussions

This section presents the results obtained from the experiments and discusses the practical issues mentioned in Section III.

### A. Experiment Overview

The results presented in the section are collected from the experiments conducted at night, to minimize the chances that external effects (human activities, co-channel interference etc) could affect the results. During the experiments all four USRP devices are located in the same lab and separated by three to five meters, and the transmit/receive antennas for each link generally have a line-of-sight signal path. The signal power transmitted by all USRP devices is approximately 0 dBm, measured via a power meter. The receiver noise figure is approximately 10 dB according to the carrier frequency and receiver amplifier gain [26] in the Labview Communications software.

### B. Observed Results and Discussions

Fig. 9 shows the channel coefficients estimated at both APs of the same OFDM symbol sent from two UEs. Each cluster of constellation points represents the channel coefficients estimated for one of the four access links shown in Fig. 2. Inside each cluster there are eight constellation points, representing the channel coefficients estimated using 8 pilot subcarriers of this OFDM symbol. For UE1 the pilot subcarriers are 11, 17, 23, 29, 36, 42, 48, 54, and for UE2 the pilot subcarriers are 12, 18, 24, 30, 37, 43, 49, 55. Subcarriers 1 to 8 and 58 to 64 are null carriers and subcarrier 33 is the DC carrier. In this figure, we can see that the pilots of each access link have similar amplitude but slight phase shift across pilot subcarriers. This is caused by the SCO mentioned in Section III, resulting from the clock drift between USRP devices, and the SCO can be obtained from (13). It can be observed that the UE signals received at AP2 have a larger phase difference across the subcarriers, indicating that there is a larger SCO between these two USRP devices. The UE signals at AP2 also have a larger amplitude variation, which could be caused by the different channel conditions or the USRP device. The packets received within the same 10-packet burst tend to have similar subcarrier phase difference and channel coefficients but with slight phase rotation. The packets received from different bursts usually have similar amplitude but a large phase difference, but the relative channel coefficients $h_{j2}/h_{j1}$ remain similar for all subcarriers if the antennas are not moved.

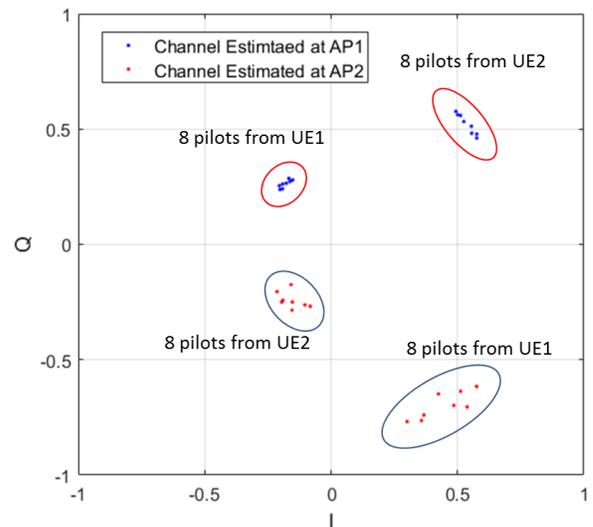

**Fig. 9 Estimated channel coefficients (a)**

Fig. 10 and Fig. 11 shows the constellation points of the superimposed UE signals received by both APs at the same time instant as the results shown in Fig. 9. The phase offsets between subcarriers are removed according to the estimated $\varphi$ in order to display a clear 16-point constellation. According to the relative channel coefficients obtained in Fig. 9, the superimposed signal that arrived at AP1 is closest to $v_{SFS}^{(3)}$ in Fig. 17, and the signal at AP2 is closest to the $v_{SFS}^{(1)}$, so the hub has selected the 11[th] mapping for the APs to encode the UE data. The complex symbols that represent the 16 combinations of the UE messages are shown in Fig. 10 and Fig. 11, and they are coded to four PNC codewords by using the selected binary mapping matrix. In this particular case the complex symbols of different UE message combinations are well separated from each other at AP1, indicating that this AP is not in any SFS and it is possible to decode both UE messages even without using the signal received from the other AP. The symbols coded to different PNC codewords are marked with different shapes of clusters and Fig. 11 shows that the symbols close to each other are coded with the same PNC codeword.

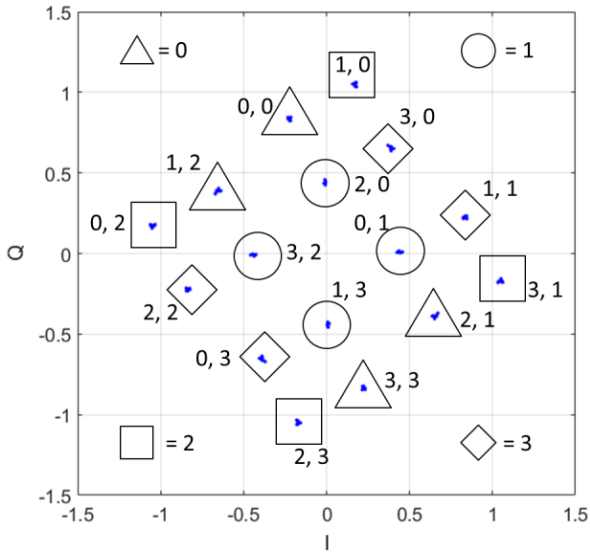

**Fig. 10 Superimposed constellation of AP1 (a)**

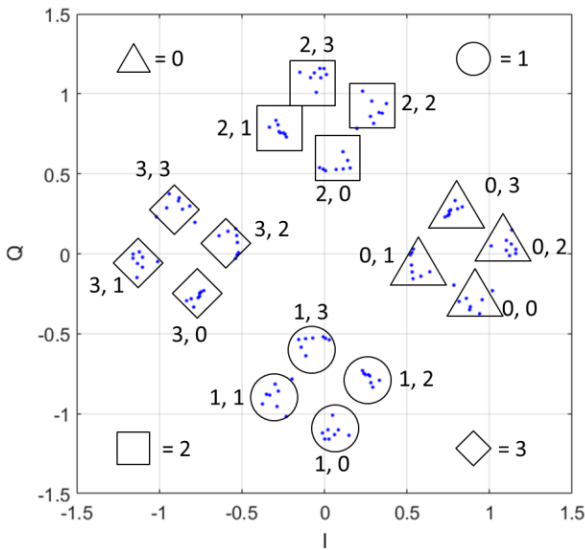

**Fig. 11 Superimposed constellation of AP2 (a)**

Fig. 12 shows the channel coefficients of another OFDM symbol received by both APs. The results are collected when the antenna is aimed differently from the case of Fig. 9. It can be seen that at this moment the sampling clock of UEs and AP2 are quite well aligned given the small phase differences between pilot subcarriers. However, the signal at AP1 has large phase differences across subcarriers, indicating roughly $t_s/4$ SCO between the two USRP devices. The maximum phase difference between two subcarriers will not exceed $\pi$ because the preamble detection keeps the SCO less than half sample duration. Although the channel coefficients of the UE signal at AP1 have quite a large phase spread across subcarriers, the relative phase between certain subcarriers remains the same. For example, the phase difference between subcarrier 11 and 12, 36 and 37, 54 and 55 are all approximately 30°, where each of these three pairs of channel coefficients are used to determine the PNC mapping of four neighbouring data carriers. In this case, all subcarriers use the same PNC mapping.

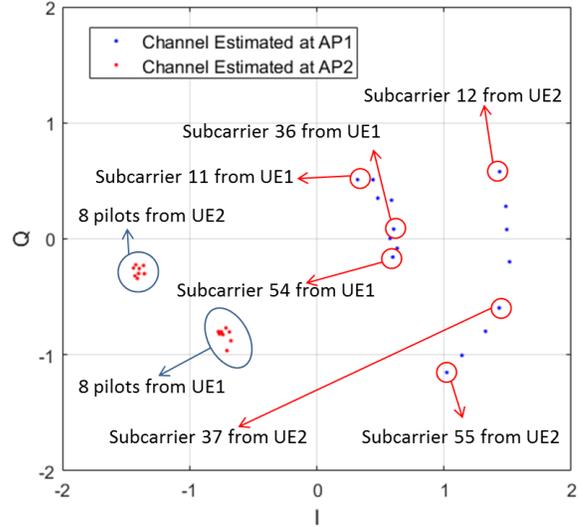

**Fig. 12 Estimated channel coefficients (b)**

Fig. 13 shows the superimposed constellation points of the UE signal that arrived at AP2 at the same time instant as the results shown in Fig. 12. According to the estimated channel coefficients in Fig. 12, the signal received by two APs are both closest to the 4[th] SFS, therefore the hub selects the 19[th] PNC mapping for the APs to encode the UE data. The symbols corresponding to the 16 combinations of UE messages are labelled in Fig. 13. Compare the results here with Fig. 10 which has 16 clearly separated clusters of complex symbols. There are 12 clusters of symbols (8 in the outer ring and 4 in the inner ring) where the four clusters in the inner ring are actually eight almost overlapped clusters. At high SNR conditions the eight inner-ring symbols will be separated so that UE data recovery using only AP2 is still possible. However, if SNR is low these eight symbols will appear as four clusters and AP2 will be in a singular fade state. Now UE data recovery must be performed at the hub with PNC coded data from both APs. In principle, the symbols close to each other should be coded to the same PNC codeword to obtain maximum distance between PNC codewords. However, for the particular channel coefficients in Fig. 12, the off-line search has to select a sub-optimal mapping for AP2 to form a full-rank and invertible mapping matrix according to rule 2) in Section II. C.

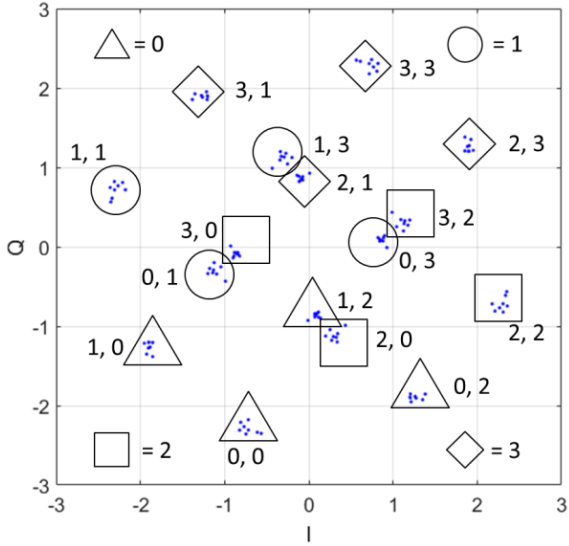

**Fig. 13 Superimposed constellation of AP2 (b)**

Fig. 14 shows the constellation of the superimposed UE signal received by AP1 at the same time instant as the signal in Fig. 13. It is clear that the symbols representing different UE message combinations are well separated from each other, so that UE data recovery is possible even without the PNC codewords from AP2. This indicates that in the rare cases where one of the APs is in an unresolvable fade state, it is still possible to recover both UE messages by using the signal received by the other AP.

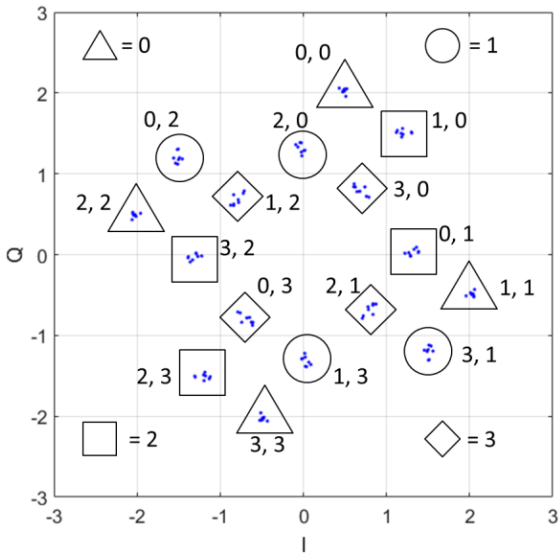

**Fig. 14 Superimposed constellation of AP1 (b)**

Fig. 15 shows the boxplot of the CFO detected at both APs for access and backhaul links before the CFO correction. The upper and lower bounds of the box represent 75% and 25% of the CFO values collected and each box has the 50% median value in the centre. The two whiskers of each box are the upper and lower 1.5 interquartile ranges (IQR). According to the 2.5 ppm accuracy of the oscillator on the USRP [21], the maximum CFO we can expect at 2.5 GHz carrier frequency is $\pm 12.5$ kHz. The maximum CFO the preamble can detect can be obtained by:

$$\Delta f_{max} = \frac{\pm \pi}{2\pi \delta t} \qquad (14)$$

where $\pm \pi$ is the maximum phase shift one short preamble can detect and $\delta t$ is the short preamble length. In our case $\delta t = 16 t_s = 16 \mu s$, so $\Delta f_{max} = 31.25$ kHz which is capable to detect the maximum CFO that could appear between any two USRP devices we use. Fig. 15 shows that the maximum CFO we have observed during the experiments is 3.7 kHz which is well within the capability of CFO detection. It can also be seen that the two USRP devices operating as AP1 and hub originally have very low clock drift.

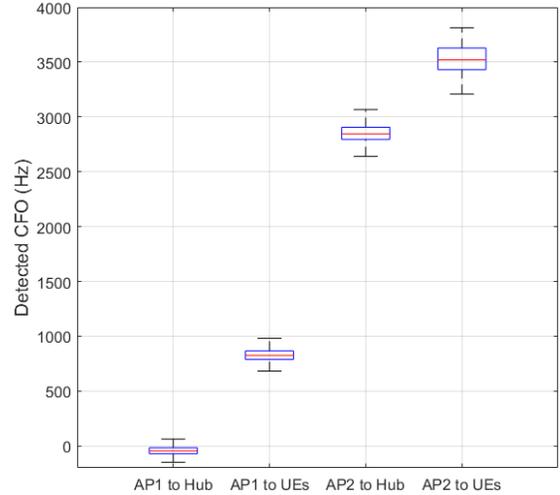

**Fig. 15 Detected CFO between USRP devices**

Fig. 16 shows the SER performance of the prototype system, compared with the simulated results in Matlab. In the simulation, the 5-node system uplink scenario shown in Fig. 2 is considered. Both UEs employ 4-QAM and the proposed PNC approach is implemented at each AP node. Perfect and estimated CSI are used to calculate the SFS in the simulation respectively, where 1 pilot symbol is employed to estimate the channel coefficients. While collecting the SER data during the experiments, the transmit antenna of two UEs are moving slowly but randomly (main beam pointing generally remains the same) in the range of a few wavelengths to evaluate the system with various channel fade states. The results show that the evaluated scheme achieves $2 \times 10^{-3}$ SER at 10 dB Eb/N0 with perfect CSI, and its SER performance is approximately 3 dB below CoMP across different Eb/N0 levels. The noise signals in the experiments are generated at the sources according to the desired Eb/N0 levels and device noise figure [26], and we can see that the SER of the prototype system follows the trend of the simulated SER quite well. The SER difference at 25 dB is because it usually takes a few minutes for the system to receive a symbol error during the experiment at high Eb/N0 levels. The SER result at 0 dB Eb/N0 is not presented because the OFDM packet detection fails with the existence of too much noise.

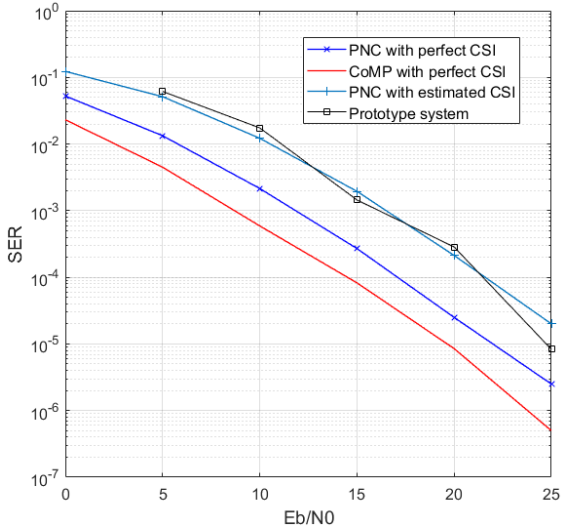

**Fig. 16 SER performance**

## C. Discussions on Scaling

An ICG can potentially involve more than 2 UEs and 2 APs to obtain further benefits on access link spectral efficiency and backhaul throughput. Compare with the direct link transmission, a 5-node ICG (as Fig. 2 shows) can increase the access link spectral efficiency by 100%. Compare with the joint reception of CoMP (one UE served by two APs) the 5-node ICG requires only 50% backhaul throughput to serve each user. After including one more UE and AP in one ICG, the access link spectral efficiency can be further increased by 50% and the backhaul throughput requirement is 67% over the 5-node ICG. However the trade-off between network performance and computational complexity must be considered. For example the superimposed constellation (as Fig. 3 shows) of three QPSK signals will have 64 constellation points with a much larger set of SFSs and PNC mappings to be produced by offline search. Using higher order modulation schemes have the similar trade off. For example in a 5-node ICG with 16-QAM, the superimposed constellation will have 256 constellation points which will result xxx SFSs.

## VI. Conclusion

In this paper the first implementation of network coded modulation which applies PNC to the uplink of two-hop networks has been presented. Applying PNC to a 5-node system has been demonstrated along with the design criteria of the binary mapping matrices. The prototype implementation using USRP testbeds has been presented in detail. The paper has highlighted several practical challenges and their solutions, including clock drift, CFO and SCO. PNC mapping selections at different channel fade states have been detailed, along with the experimental results arising from their use. The SER results show that the prototype system closely matches results obtained from simulation. Thus, PNC can achieve much better backhaul efficiency compared with CoMP with minimal performance degradation, meaning that it can bring significant benefits to the development of ultra-dense 5G and beyond wireless systems.

## VII. Appendix

### A. SFS from Off-line Search

The SFSs for the 4-QAM modulation are expressed in the form of equation (8), assuming $h_{j1} = 1 + 0i$. Five SFSs are found by the Offline-search, the values are provided below as well as the associated superimposed constellations.

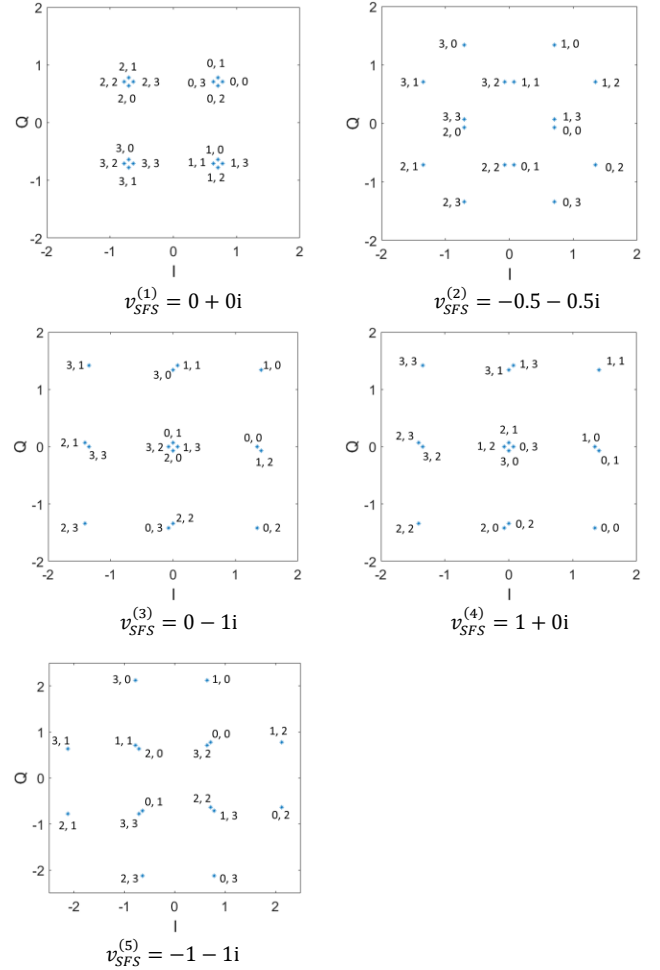

**Fig. 17 SFSs and constellations**

### B. Mapping Matrices from Off-line Search

The binary mapping matrices are given by $\mathbf{M}_{ij}$ where $i$ is the SFS at AP1 and $j$ is the SFS at AP2.

$$\mathbf{M}_{11} = \begin{bmatrix} 0 & 1 & 0 & 0 \\ 1 & 0 & 0 & 0 \\ 0 & 0 & 0 & 1 \\ 0 & 0 & 1 & 0 \end{bmatrix} \quad \mathbf{M}_{12} = \begin{bmatrix} 0 & 1 & 0 & 0 \\ 1 & 0 & 0 & 0 \\ 0 & 0 & 1 & 1 \\ 1 & 1 & 1 & 0 \end{bmatrix}$$

$$\mathbf{M}_{13} = \begin{bmatrix} 0 & 1 & 0 & 0 \\ 1 & 0 & 0 & 0 \\ 1 & 0 & 0 & 1 \\ 0 & 1 & 1 & 0 \end{bmatrix} \quad \mathbf{M}_{14} = \begin{bmatrix} 0 & 1 & 0 & 0 \\ 1 & 0 & 0 & 0 \\ 0 & 1 & 0 & 1 \\ 1 & 0 & 1 & 0 \end{bmatrix}$$

$$\mathbf{M}_{15} = \begin{bmatrix} 0 & 1 & 0 & 0 \\ 1 & 0 & 0 & 0 \\ 0 & 1 & 1 & 1 \\ 0 & 0 & 1 & 0 \end{bmatrix} \quad \mathbf{M}_{21} = \begin{bmatrix} 0 & 0 & 1 & 1 \\ 1 & 1 & 0 & 1 \\ 1 & 0 & 0 & 0 \\ 0 & 1 & 0 & 0 \end{bmatrix}$$

$$\mathbf{M}_{22} = \begin{bmatrix} 0 & 1 & 0 & 0 \\ 1 & 0 & 0 & 0 \\ 0 & 0 & 0 & 1 \\ 0 & 0 & 1 & 0 \end{bmatrix} \quad \mathbf{M}_{23} = \begin{bmatrix} 0 & 0 & 1 & 1 \\ 1 & 1 & 0 & 1 \\ 1 & 0 & 0 & 1 \\ 0 & 1 & 1 & 0 \end{bmatrix}$$

$$\mathbf{M}_{24} = \begin{bmatrix} 0 & 0 & 1 & 1 \\ 1 & 1 & 0 & 1 \\ 1 & 0 & 1 & 0 \\ 0 & 1 & 0 & 1 \end{bmatrix} \quad \mathbf{M}_{25} = \begin{bmatrix} 0 & 0 & 1 & 1 \\ 1 & 1 & 0 & 1 \\ 1 & 0 & 1 & 1 \\ 0 & 1 & 1 & 1 \end{bmatrix}$$

$$\mathbf{M}_{31} = \begin{bmatrix} 0 & 1 & 1 & 0 \\ 1 & 0 & 0 & 1 \\ 1 & 0 & 0 & 0 \\ 0 & 1 & 0 & 0 \end{bmatrix} \quad \mathbf{M}_{32} = \begin{bmatrix} 0 & 1 & 1 & 0 \\ 1 & 0 & 0 & 1 \\ 0 & 0 & 1 & 1 \\ 1 & 1 & 1 & 0 \end{bmatrix}$$

$$\mathbf{M}_{33} = \begin{bmatrix} 0 & 1 & 1 & 0 \\ 1 & 0 & 0 & 1 \\ 0 & 1 & 0 & 0 \\ 0 & 0 & 0 & 1 \end{bmatrix} \quad \mathbf{M}_{34} = \begin{bmatrix} 0 & 1 & 1 & 0 \\ 1 & 0 & 0 & 1 \\ 0 & 1 & 0 & 1 \\ 0 & 0 & 0 & 1 \end{bmatrix}$$

$$\mathbf{M}_{35} = \begin{bmatrix} 0 & 1 & 1 & 0 \\ 1 & 0 & 0 & 1 \\ 0 & 1 & 1 & 1 \\ 1 & 0 & 1 & 1 \end{bmatrix} \quad \mathbf{M}_{41} = \begin{bmatrix} 0 & 1 & 0 & 1 \\ 1 & 0 & 1 & 0 \\ 0 & 1 & 0 & 0 \\ 1 & 0 & 0 & 0 \end{bmatrix}$$

$$\mathbf{M}_{42} = \begin{bmatrix} 0 & 1 & 0 & 1 \\ 1 & 0 & 1 & 0 \\ 0 & 0 & 1 & 1 \\ 1 & 1 & 1 & 0 \end{bmatrix} \quad \mathbf{M}_{43} = \begin{bmatrix} 0 & 1 & 0 & 1 \\ 1 & 0 & 1 & 0 \\ 0 & 1 & 1 & 0 \\ 0 & 1 & 0 & 0 \end{bmatrix}$$

$$\mathbf{M}_{44} = \begin{bmatrix} 0 & 1 & 0 & 1 \\ 1 & 0 & 1 & 0 \\ 1 & 0 & 0 & 0 \\ 0 & 0 & 0 & 1 \end{bmatrix} \quad \mathbf{M}_{45} = \begin{bmatrix} 0 & 1 & 0 & 1 \\ 1 & 0 & 1 & 0 \\ 0 & 1 & 1 & 1 \\ 1 & 1 & 0 & 0 \end{bmatrix}$$

$$\mathbf{M}_{51} = \begin{bmatrix} 0 & 1 & 1 & 1 \\ 1 & 0 & 1 & 1 \\ 0 & 1 & 0 & 0 \\ 0 & 0 & 0 & 1 \end{bmatrix} \quad \mathbf{M}_{52} = \begin{bmatrix} 0 & 1 & 1 & 1 \\ 1 & 0 & 1 & 1 \\ 0 & 0 & 1 & 1 \\ 1 & 1 & 1 & 0 \end{bmatrix}$$

$$\mathbf{M}_{53} = \begin{bmatrix} 0 & 1 & 1 & 1 \\ 1 & 0 & 1 & 1 \\ 0 & 1 & 1 & 0 \\ 1 & 0 & 0 & 1 \end{bmatrix} \quad \mathbf{M}_{54} = \begin{bmatrix} 0 & 1 & 1 & 1 \\ 1 & 0 & 1 & 1 \\ 1 & 0 & 1 & 0 \\ 0 & 1 & 0 & 0 \end{bmatrix}$$

$$\mathbf{M}_{55} = \begin{bmatrix} 0 & 1 & 1 & 1 \\ 1 & 0 & 1 & 1 \\ 0 & 1 & 0 & 0 \\ 0 & 0 & 0 & 1 \end{bmatrix}$$